\documentclass{ifacconf}

\usepackage{graphicx}      
\usepackage{natbib}        

\usepackage{amsmath,amssymb,amsfonts}
\usepackage{algorithm}
\usepackage{algpseudocode}
\usepackage{tikz}

\usepackage{xcolor}

\usepackage{mathtools}
\usepackage{bm}
\usepackage{bbm}

\usepackage{graphicx}
\usepackage{amssymb}
\usepackage{xcolor}

\usepackage{algorithm}
\usepackage{algpseudocode}

\newcommand{\BR}{\mathbb{R}} 
\newcommand{\R}{\mathbb{R}} 
\newcommand{\BN}{\mathbb{N}} 
\newcommand{\BC}{\mathbb{C}} %

\newcommand{\bI}{\mathbf{I}} %
\newcommand{\bZero}{\mathbf{0}} %

\newcommand{\CR}{\mathcal{R}} %
\newcommand{\CC}{\mathcal{C}} %

\newcommand{\Ker}{{\rm Ker}}

\newcommand{\Span}{\rm Span} 

\newcommand{\Rank}{\text{rank}} %

\newcommand*{\QE}{\hfill\ensuremath{\blacksquare}}	

\newcommand*{\QEDA}{\null \hfill\ensuremath{\triangle}}
\newcommand*{\SQ}{\hfill\ensuremath{\square}}

\begin{document}
\begin{frontmatter}

\title{Control of Discrete-Time Linear Systems with Charge-Balanced Inputs\thanksref{footnoteinfo}} 

\thanks[footnoteinfo]{This publication was part
of the project Dutch Brain Interface Initiative (DBI2) with project
number 024.005.022 of the research programme Gravitation, which
was financed by the Dutch Ministry of Education, Culture and
Science (OCW) via the Dutch Research Council (NWO).}

\author[First]{Yuzhen Qin} 
\author[Second]{Zonglin Liu} 
\author[First]{Marcel van Gerven}

\address[First]{Department of Machine Learning and Neural Computing, Donders Institute for Brain, Cognition and Behaviour, Radboud University, Nijmegen, 6525XZ, the Netherlands (e-mail: \{yuzhen.qin, m.vangerven\}@donders.ru.nl)}
\address[Second]{Control and System
Theory Group, Department of Electrical Engineering and Computer Science, University of Kassel, 34121, Kassel,
Germany (e-mail: z.liu@uni-kassel.de)}

\begin{abstract}
Electrical brain stimulation relies on externally applied currents to modulate neural activity, but safety constraints require each stimulation cycle to be charge-balanced, enforcing a zero net injected charge. However, how such charge-balanced stimulation works remains poorly understood. This paper investigates the ability of charge-balanced inputs to steer state trajectories in discrete-time linear systems. Motivated by both open-loop and adaptive neurostimulation protocols, we study two practically relevant input structures: periodic (repetitive) charge-balanced inputs and non-repetitive charge-balanced inputs. For each case, we derive novel reachability and controllability conditions. The theoretical results are further validated through numerical demonstrations of minimum-energy control input design.

 
\end{abstract}


\end{frontmatter}

\section{Introduction}

\begin{figure}[t]
	\centering
	\begin{tikzpicture}
		\node at(-1,0.1) {\includegraphics[scale=0.8]{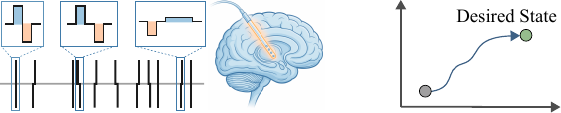}};		
        \node[scale=0.8] at(-5.1,0.7) {$(a)$};
        \node[scale=0.8] at(0.3,0.7) {$(b)$};
	\end{tikzpicture}
    \vspace{-8pt}
	\caption{Charge-balanced pulses steer brain states.}		
	\label{fig_motivation}
    \vspace{-2pt}
\end{figure}

Electrical brain stimulation has emerged as a key tool for treating neurological and psychiatric disorders, including Parkinson’s disease, epilepsy, chronic pain, and depression \citep{krauss2021technology,bergey2015long}. Clinical systems such as deep brain stimulation (DBS), cortical stimulation, and vagus nerve stimulation (VNS) all rely on injected currents to modulate the activity of neural populations or peripheral nerves. A central safety requirement in these devices is that stimulation at the electrode–tissue interface is charge-balanced: the net injected charge over each stimulation cycle must be (approximately) zero to avoid irreversible electrochemical reactions, tissue damage, and electrode corrosion  \citep{ng2024biophysical}, which is typically enforced via biphasic, charge-balanced pulses (Fig.~\ref{fig_motivation} (a)).

Despite clinical success, the mechanisms by which electrical stimulation produces therapeutic benefit remain ill-understood, limiting our ability to predict, optimize, and personalize stimulation strategies. There has been increasing interest in computational modeling of neural dynamics under electrical stimulation \citep{farooqi2024deep}. The fundamental question becomes whether the stimulation input can steer the neural system from its current state toward a desired therapeutic state (e.g., desynchronized activity in Parkinsonian circuits, seizure termination in epilepsy, illustrated in Fig.~\ref{fig_motivation} (b)). Understanding the controllability and reachability of these models, and the extent to which stimulation constraints limit them, is therefore central to designing principled neurostimulation protocols.
Motivated by this perspective, this paper studies linear systems controlled by brain-stimulation-type inputs that are charge-balanced. Linearized models of neural activity are commonly used to approximate local dynamics around operating points or to capture dominant oscillatory modes. Our goal is to characterize how the charge-balanced constraint affects reachability and controllability.

\textbf{Related work.} 
 Charge-balanced inputs have been studied for controlling spiking neurons \citep{dasanayake2015constrained, dasanayake2012charge} and for desynchronizing coupled neural oscillators \citep{mau2022optimizing, moehlis2025nearly}. Recent model-based methods on stimulation take practical inputs into account~\citep{tamekue2025control, olumuyiwa2025proportional}. Linear systems with constrained inputs have been extensively studied. Recent advances address a range of constraints, including input saturation and general polyhedral bounds, for various problems such as MPC and online optimization \citep{porcari2024data, govindaraj2023saturated, liu2024stability, nonhoff2023online}. However, aside from a few studies that employ sinusoidal inputs for stabilization \citep{qin2025vibrational}, charge-balanced input constraints remain largely unexplored, leaving fundamental questions open and motivating the present work.
 
\newpage
\textbf{Contribution.} To the best of our knowledge, this work is among the first to explicitly incorporate charge-balanced inputs into the formal analysis of control systems. We consider settings motivated by both open-loop and adaptive neurostimulation, examining repetitive and non-repetitive charge-balanced input structures. We derive sufficient conditions for reachability and controllability that clarify when charge balance reduces effective input directions and when full-state steering can be recovered through temporal structure. In addition, we provide closed-form minimum-energy charge-balanced control laws and demonstrate their performance through numerical examples.

\section{Problem Formulation}\label{prom_formu}

Consider a discrete-time linear time-invariant system described by 
\begin{align}\label{main}
    &x_{k+1} = A x_{k} +Bu_{k},&k=0,1, 2, \cdots,
\end{align}
where $x_k\in \R^n, u_k\in\R^m$. We impose the following assumption to ensure that the control inputs satisfy the \textit{charge-balanced condition} commonly required in electrical brain stimulation paradigms. 
\begin{assum}\label{charge-balance}
    Given an integer $h >0$,  the control inputs satisfy $
       \sum_{i=0}^{h-1}u_{ph+i} =\bZero$ for all $p=0,1,2,\cdots$.
\end{assum}
This assumption states that within each input block, motivated by the notation of a stimulation cycle in brain stimulation, the positive and negative inputs delivered to the system are balanced. The block length $h$ is a designable parameter. Given that brain stimulation inputs generally require charge balancing within a very short time window, it is desirable to choose $h$ to be small.

For notational simplicity, we write the inputs in each block into a vector $U_p \coloneqq [u_{ph}^\top \;\; u_{ph+1}^\top\;\;\cdots \;\;u_{(p+1)h-1}^\top]^\top \in \R^{mh}$ for $p=0,1,2, \cdots$. 
Define $R \coloneqq \mathbf{1}_h^\top \otimes  \bI_m \in \BR^{m\times mh}$. Then, Assumption~\ref{charge-balance} can be rewritten as
\begin{align}\label{input_constrain}
    &RU_p = \bZero, &\forall p=0, 1, 2, \cdots.
\end{align}

In this paper, we investigate whether and how desired states can be reached with the control inputs constrained to satisfy the charge-balanced condition~\eqref{input_constrain}. Next, we define reachability and  controllability.

\begin{defn}
    For the system~\eqref{main}, a target state $x_f\in\R^n$ is said to be \emph{reachable under charge-balanced inputs} from an initial condition $x_0\in\R^n$ if there exist finite integers $h\ge 2, b\ge 1$, and a sequence of control inputs $U_0,U_1,\dots, U_{b-1}$, each satisfying~\eqref{input_constrain}, such that $x_{bh}=x_f$. The system is said to be \emph{controllable under charge-balanced inputs} if any $x_f$ is reachable under charge-balanced inputs from any $x_0$.
\end{defn}

Here, reachability is defined at the boundary of the terminal block to avoid cases in which the terminal state occurs within a block. This ensures that the overall sequence of applied inputs remains charge-balanced.  

The objectives of this papers are twofold: (i) to derive conditions for reachability and controllability, and (ii) to design minimum-energy control laws for steering the system to specified target states.

Given a block length $h$ and the block horizon $b$, the minimum-energy control problem for the system~\eqref{main} with input constraint~\eqref{input_constrain} is defined by:
\begin{align}\label{min_control}
    \min_{U_p,p=0,\dots,b-1} \hspace{20pt}&J:=\sum_{p=0}^{n-1}\|U_p\|^2_2, \nonumber\\
    \text{s.t.} \hspace{20pt}& x_{k+1} = A x_{k} +Bu_{k},\\
        &  x_0=x_0, x_{N}=x_f\nonumber,\\
        & 
        RU_p = \bZero, \forall p=0,1,2,\dots,b-1. \nonumber
\end{align}

Concerning these objectives, we  consider two situations. 

First, we will consider the general case, where charge-balanced blocks are non-repetitive, that is,  $U_p$ and $U_q$ are allowed to be different for any $p,q$. The results in this setting provide theoretical groundwork for future adaptive control strategies that could guide the design of brain-stimulation protocols responsive to neural dynamics. 

Second, we examine a special case in which all charge-balanced blocks are identical, i.e, $U_0=U_1=\cdots$.  The results here offer interpretation and theoretical justification for open-loop control strategies based on constant, pre-programmed stimulation patterns, which are mostly used in current clinical and experimental paradigms. 

\section{Non-repetitive blocks}

To proceed, let us derive a lifted system based on charge-balanced blocks. Recall that each block satisfies~\eqref{input_constrain}. Now, define the matrix $Q\in\R^{mh\times m(h-1)}$ with columns forming the basis of $\Ker(R)$. Without loss of generality, one can select $Q$ with orthonormal columns. Then, it holds that $Q^\top Q=I_{m(h-1)}$. Consequently, it follows that control inputs within each charge-balanced block $p$ satisfy
\begin{align}\label{def_Q}
    \exists w_p\in\R^{m(h-1)}: U_p=Qw_p.
\end{align}

Denote 
\begin{equation}\label{def_S}
    S:= \begin{bmatrix}
        A^{h-1}B& A^{h-2}B& \cdots & AB & B
    \end{bmatrix}\in \R^{n\times mh}.
\end{equation}
Then, at each block boundary, it holds that
\begin{equation}\label{lifted_1}
    x_{(p+1)h} = A^h x_{ph}+SU_p =A^h x_{ph}+SQw_p.
\end{equation}

\begin{defn}
    For the system~\eqref{lifted_1}, a target state $x_f\in\R^n$ is said to be \emph{reachable} from an initial condition $x_0\in\R^n$ if there exist finite integers $h\ge 2, b\ge 1$, and a sequence of inputs $w_0,w_1,\dots, w_{b-1}$ such that $x_{bh}=x_f$. The system is said to be \emph{controllable} if any $x_f$ is reachable from any $x_0$.
\end{defn}

\begin{lem}\label{connect_reach_cntr}
    The system~\eqref{main} is reachable under charge-balanced inputs if the system~\eqref{lifted_1} is reachable. Further, the system~\eqref{main} is controllable under charge-balanced inputs if the system~\eqref{lifted_1} is controllable (i.e.,  controllable $(A^h,SQ)$). 
\end{lem}

From this lemma, studying the reachability and controllability of the original system~\eqref{main} reduces to the exploration of those of the lifted system~\eqref{lifted_1}. 
For notational simplicity, let $\bar A \coloneqq A^h, \bar B\coloneqq SQ$. Then, we have the block-to-block lifted system given by
\begin{equation}\label{lifted}
    x_{(p+1)h} = \bar A x_{ph}+\bar B w_p.
\end{equation}

\subsection{Reachability and minimum-energy control}

Similar to classic linear systems theory, we can define the $b$-block reachability matrix for the lifted system~\eqref{lifted}:
\begin{equation}
    \CR_b \coloneqq [\bar A^{b-1}_h \bar B\;\;\bar A^{b-2}\bar B\;\;\cdots\;  \;\bar B].
\end{equation}
Then, the $b$-block controllability Gramian is defined by
\begin{equation}
    G_{h,b}\coloneqq \CR_b \CR_b ^\top =\sum_{p=0}^{b-1} \bar A^p \bar B \bar B^\top (\bar A^\top)^p.
\end{equation}

Observe that 
\begin{equation*}
    \|U_p\|_2^2=\sum_{p=0}^{b-1} w_p^\top (Q^\top Q)w_p=\sum_{p=0}^{b-1} \|w_p\|_2^2
\end{equation*}
where the last equality has used the fact $Q^\top Q=I$. Then, the minimum energy problem~\eqref{min_control} reduces to
\begin{align}\label{min_control_lifted}
    \min_{w_p,p=0,\dots,b-1} \hspace{20pt}&\sum_{p=0}^{n-1}\|w_p\|^2_2, \nonumber\\
    \text{s.t.} \hspace{20pt}& x_{(p+1)h} = \bar A x_{ph} +\bar B w_{p},\\
        &  x_0=x_0, x_{bh}=x_f,\nonumber
\end{align}
as minimizing $\sum_{p=0}^{n-1}\|w_p\|^2_2$ is equavilent to minimizing $\|U_p\|_2^2$. Then, solving the problem~\eqref{min_control_lifted} for  $w_p^*$ immediately yields the desired solution $U_p^*=Qw_p^*$ to~\eqref{min_control}.  
The lemma below directly follows the classic results on reachability of linear systems \citep{kailath1980linear}. 

\begin{lem}
    Given an initial condition $x_0\in\R^n$ and a target state $x_f\in\R^n$, if $x_f-\bar A^b x_0 \in \Span (G_{h,b})$, $x_f$ is reachable from $x_0$ in $b$ blocks in the system~\eqref{main}. The unique minimum-energy control inputs that drive the system from $x_0$ to $x_f$ are given by 
    \begin{align*}
    U_p^* =Qw_p^*= QQ^\top S^\top (\bar A^\top)^{b-1-p}G_{h,b}^{\dagger} (x_f-\bar A^bx_0),
\end{align*}
for $p=0,1,\dots,b-1$, where $(\cdot)^\dagger$ denotes the  Moore–Penrose pseudoinverse. 
\end{lem}

Further, if the $n$-block controllability Gramian satisfies $\Rank(G_{h,n})=n$ for some $h\ge 2$, the system~\eqref{lifted} is controllable, which implies that the original system~\eqref{main} is controllable under charge-balanced inputs.

A natural question arises: What conditions does the original system~\eqref{main} need to satisfy so that it is controllable under charge-balanced inputs?

\subsection{Controllability}
Below, we present the main result in this section. 

\begin{thm}[Sufficient condition]\label{generic_ctr}
    The system~\eqref{main} is controllable under charge-balanced inputs if 
    \begin{enumerate}
        \item[(i)] the pair $(A, B)$ is controllable,
        \item[(ii)] $1\notin \sigma(A)$, i.e., $A$ has no eigenvalue equal to $1$, and
        \item[(iii)] there exists $h\ge 2$ such that $A^h$ has a simple spectrum, i.e., all eigenvalues of $A^h$ are distinct.  
    \end{enumerate}
\end{thm}

\begin{pf}
    From Lemma~\ref{connect_reach_cntr}, it suffices to show the pair $(\bar A, \bar B)$ is controllable. We prove this by employing the Popov–Belevitch–Hautus (PBH) controllability test. 
    
    By assumption (iii), $\bar A=A^h$ is diagonalizable with distinct eigenvalues. Therefore, $A$ and $A^h$ are simultaneously diagonalizable. In other words, all left eigenvectors of $A^h$ are also left eigenvectors of $A$. Now, let $\phi^\top$ be a left eigenvector of $\bar A$ with eigenvalue $\mu$, i.e., $\phi^\top \bar A = \mu \phi^\top$. 
Then, it follows that
\begin{equation*}
    \phi^\top A = \lambda \phi^\top
\end{equation*}
for some $\lambda\in\sigma(A)$ satisfying $\mu=\lambda^h$.

For such a pair $(\phi^\top,\lambda)$, using the identity $\phi^\top A^k=\lambda^k\phi^\top$, one can derive that
\begin{align*}
\phi^\top S
&= \begin{bmatrix}
    \phi^\top A^{h-1} B,& \phi^\top A^{h-2} B, & \cdots &\phi^\top B
\end{bmatrix}\\
&=\begin{bmatrix}
    \lambda^{h-1}\phi^\top B,& \lambda^{h-2}\phi^\top B, & \cdots &\phi^\top B
\end{bmatrix} = r_h(\lambda)\otimes (\phi^\top B),
\end{align*}
where $r_h(\lambda) \coloneqq [\lambda^{h-1},\dots,1]$.

We now characterize $\phi^\top \bar B=\phi^\top S Q$.
Since $Q$ has columns spanning $\Ker(R)$, for $v\in\BC^{mh}$, $v^\top Q=0$ holds if and only if 
\begin{align*}
    v \in \Ker(Q^\top) &\Longleftrightarrow  v \in \big( \Span (Q) \big)^ \perp \\
    & \Longleftrightarrow  v \in \big(\Ker(R) \big)^ \perp \Longleftrightarrow v\in \Span (R^\top).
\end{align*}
As $R = \mathbf{1}_h^\top \otimes  \bI_m$, it follows that
\begin{equation*}
    \Span (R^\top) = \{ \begin{bmatrix}
        \alpha^\top & \alpha^\top & \cdots & \alpha^\top
    \end{bmatrix}^\top \in \BC^{mh}: \alpha \in \BC^{m}  \}
\end{equation*}

Therefore, $\phi^\top S Q=0$ if and only if the $h$ blocks of $\phi^\top S$ are identical.
Since these blocks are
\begin{equation*}
v_r = \lambda^{h-r}\phi^\top B,\qquad r=1,\dots,h,
\end{equation*}
they are equal if and only if either
\begin{equation*}
\phi^\top B = 0,
\qquad\text{or}\qquad
\lambda^{h-1}=\lambda^{h-2}=\cdots=1,
\end{equation*}
the latter implying $\lambda=1$.

By assumption (i), $(A,B)$ is controllable, hence by the PBH test,  
$\phi^\top B\neq 0$ for every left eigenpair $(\phi^\top,\lambda)$ of $A$.
By assumption (ii), $\lambda\neq 1$ for all eigenvalues of $A$.
Thus neither of the two conditions above can hold.
Consequently, $\phi^\top S Q \neq 0$ for every left eigenpair  $(\phi^\top,\mu)$ of $A^h$. 
By PBH, $(A^h,SQ)$, i.e., $(\bar A, \bar B)$, is controllable.\SQ
\end{pf}

Following the proof of Theorem~\ref{generic_ctr}, it can be observed that if $\lambda=1$ is an eigenvalue of $A$, it is also an eigenvalue of $\bar A=A^h$ satisfying $\phi^\top \bar A=\phi^\top$. Yet, one can derive that $\phi^\top \bar B=\phi^\top SQ = 0$, indicating that $(\bar A,\bar B)$ is not controllable. Therefore, the corollary below follows.  
\begin{cor}
    The system~\eqref{main} is controllable under charge-balanced inputs \textit{only if}: (i) the pair $(A, B)$ is controllable, and (ii)  $1\notin \sigma(A)$, i.e., $A$ has no eigenvalue equal to $1$.
\end{cor}

Together with Theorem~\ref{generic_ctr}, this corollary yields straightforward criteria for determining whether a system is controllable under charge-balanced inputs. Since conditions (i) and (ii) are necessary, one only needs to verify condition (iii). Note that condition (iii) depends on the block length $h$. With an additional assumption, the following lemma provides a way to guarantee controllability.

\begin{lem}\label{h_exist}
    Assume that the pair $(A,B)$ is controllable, $1\notin \sigma(A)$, and $A$ has distinct eigenvalues. Then, there always exists an integer $h\ge 2$ such that the pair $(\bar A, \bar B)$ is controllable. 
\end{lem}
\begin{pf}To construct the proof, it suffices to show the eigenvalues of $\bar A$ are also distinct. Let the eigenvalues of $A$ be $\lambda_1,\lambda_2,\dots, \lambda_n$. Then, the eigenvalues of $\bar A =A^h$ are 
\begin{align*}
    \lambda_1^h,\lambda_2^h,\dots, \lambda_n^h.
\end{align*}
If $ \lambda_i=0$ for some $i$, $\lambda_i^h\neq \lambda_j^h$ for any other $j$. Next, let us consider any pair of non-zero eigenvalues of $A$, $\lambda_i$ and $\lambda_j$. It holds that $\lambda_i^h = \lambda_j^h$ if and only if there exist two positive integers $k_{ij}$ and $\ell_{ij}$ such that 
\begin{align}\label{ratio_root}
    &r_{ij}\coloneqq \frac{\lambda_i}{\lambda_j} = \exp\left({\frac{2\pi k_{ij}\eta}{\ell_{ij}}} \right), &\text{ where } \eta^2=-1 ,
\end{align}
i.e., the ratio $r_{ij}$ is a root of unity. For any pair of $(i,j)$ such that $r_{ij}$ is a root of unity, define
\begin{equation*}
    m_{ij} \coloneqq \min \{k\in \BN: r_{ij}^k=1\},
\end{equation*}
which is called the order of $r_{ij}$. 
Now, choose 
\begin{equation}\label{sel_h}
    h=\text{lcm} \{ m_{ij}:i\neq j \}+1,
\end{equation}
where $\rm{lcm}\{\cdot\}$ denotes the least common multiple. This $h$ ensures that it is not divisible by any $m_{ij}$. As a result,
\begin{equation*}
    \left( \frac{\lambda_i}{\lambda_j}\right)^h \neq 1,
\end{equation*}
implying that all the eigenvalues of $\bar A$ are distinct, which completes the proof. \SQ
\end{pf}

Lemma~\ref{h_exist} ensures that, under stated conditions, one can carefully select the length $h$ of each charge-balanced block so that the resulting system is controllable. It is worth noting that, although \eqref{sel_h} provides a sufficient condition for selecting $h$, it is conservative. In practice, often a much smaller $h$ suffices. The next example illustrates this point.

\begin{exmp}\label{exmp_1}
    Consider the system matrices 
    \begin{align*}
        &A=\begin{bmatrix}
            -{1}/{2} \;\;& -{\sqrt{3}}/{2} \\
            {\sqrt{3}}/{2} \;\;& -{1}/{2}
        \end{bmatrix}, & B=\begin{bmatrix}
            1\\0
        \end{bmatrix}.
    \end{align*}
     It can be verified that the pair $(A,B)$ is controllable. The eigenvalues of $A$ are $\lambda_{1,2}=e^{\pm2\pi\eta/3}$ with $\eta^2=-1$, which are the third roots of unity. According to \eqref{sel_h}, selecting $h=4$ ensures that the pair $(\bar A, \bar B)$ is controllable. However, we next show that $h=2$ is already sufficient. 

     When $h=2$, the charge-balanced constraint~\eqref{input_constrain} implies that $R=[1\;\;1]$. Then, one can select $Q=[1\;{-1}]^\top/\sqrt{2}$. Subsequently, it can be calculated that
     \begin{align*}
         SQ= [AB\;\;B]Q=\frac{1}{\sqrt{2}}\begin{bmatrix}
             -{1}/{2}\;\;&1\\
            \sqrt{3}/{2}\;\;& 0
         \end{bmatrix}\begin{bmatrix}
             1\\-1
         \end{bmatrix}=\begin{bmatrix}
             -{3\sqrt{2}}/{4}\;\\ \sqrt{6}/{4}
         \end{bmatrix}.
     \end{align*}
     To check the controllability of the pair $(\bar A,SQ)=(A^2,\bar B)$, we calculate the controllability matrix
     \begin{align*}
         \CC(A^2,SQ)=\begin{bmatrix}
             -{3\sqrt{2}}/{4}\;\;&{3\sqrt{2}}/{4}\\
           \sqrt{6}/{4}\;\;& \sqrt{6}/{4}
         \end{bmatrix},
     \end{align*}
     As $\Rank(\CC(A^2,SQ))=2$, $(\bar A,SQ)$ is controllable.  \QEDA
\end{exmp}

For the system matrix $A$ that has all real eigenvalues, the following corollary provides a universal way to choose $h$.

\begin{cor}
    Assume that $(A,B)$ is controllable, $1\notin \sigma(A)$, and all the eigenvalues of $A$ are \textit{real} and distinct. Then, the pair $(\bar A, SQ)$ is controllable when $h=3$. 
\end{cor}
The corollary follows from the fact that if distinct real eigenvalues of $A$ satisfy $\lambda_1\neq \lambda_2$, then $\lambda_1^3\neq \lambda_2^3$. 

\begin{cor}\label{non_rep_control}
    Assume the conditions in Theorem~\ref{generic_ctr} are satisfied. Then,  the unique minimum-energy solution to~\eqref{min_control} is 
$
    U_p = QQ^\top S^\top (\bar A^\top)^{b-1-p}G_{h,b}^{-1} (x_f-\bar A^bx_0).
$   
\end{cor}

\begin{figure}[t]
	\centering
	\begin{tikzpicture}
		\node at(0,0) {\includegraphics[scale=0.6]{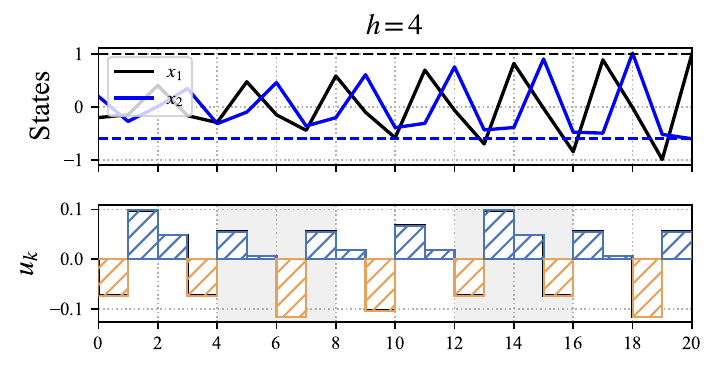}};
        \node at(0,-3.8) {\includegraphics[scale=0.6]{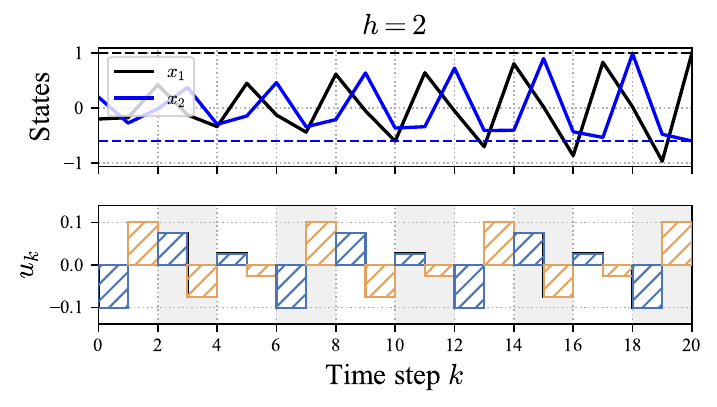}};
		
	\end{tikzpicture}
	\vspace{-14pt}
	\caption{The minimum-energy control inputs steer the system to a target state (dashed lines) in $b=5$ or $b=10$ charge-balanced blocks.}	
	\label{non_repetitive_minimum}
\end{figure}

\begin{exmp}
    For the system in Example~\ref{exmp_1}, suppose that the goal is to steer the system from $x_0=[{-0.2}\;0.2]^\top$ to $x_f=[1\;{-0.6}]^\top$ in $N=20$ steps. For both $h=4$ and $h=2$, the minimum-energy control inputs can be computed using Corollary~\ref{non_rep_control}, which, as shown in Fig.~\ref{non_repetitive_minimum}, effectively achieve the desired steering. Notice that the oscillatory trajectories result from the nature of charge-balanced control---alternating positive and negative inputs. 
\end{exmp}

\begin{rem}
    Note that the condition in Theorem~\ref{generic_ctr}, particularly (iii),  provides only sufficient guarantees for controllability. For instance, one can select $h=3$ for Example~\ref{exmp_1}. Although $A^3$ has two identical eigenvalues, violating the theorem’s hypotheses, the lifted pair $(A^3,SQ)$ is still controllable. Characterizing necessary and sufficient conditions remains an open research question.
\end{rem}

\section{Repetitive blocks}
Currently, the most widely used brain stimulation approaches for treating neurological disorders such as Parkinson’s disease are still open-loop. These methods deliver charge-balanced pulse trains with predetermined, fixed parameters to the brain tissue. To provide new insights into how such fixed stimulation patterns influence brain dynamics, we assume that all charge-balanced control blocks in \eqref{input_constrain} are identical, i.e., $U_p=U_q$ for all $p, q$, leading to periodic block inputs. 

This is a special case of the non-repetitive situation. It follows from~\eqref{def_Q} that there exists $w\in\R^{m(h-1)}$ such that $U=Qw$. Consequently, the block-to-block lifted system becomes 
\begin{align}\label{lifted_repetitive}
    x_{(p+1)h} = \bar A x_{ph} + \bar Bw,
\end{align}
where $\bar A =A^h$ and $\bar B= SQ$. 

\begin{defn}
    For the system~\eqref{lifted_repetitive}, a target state $x_f\in\R^n$ is said to be \emph{reachable} from an initial condition $x_0\in\R^n$ if there exist finite integers $h\ge 2, b\ge 1$ and inputs $w\in \R^{m(h-1)}$ such that $x_{bh}=x_f$. The system is said to be \emph{controllable} if any $x_f$ is reachable from any $x_0$.
\end{defn}

Similar to Lemma~\ref{connect_reach_cntr}, the system~\eqref{main} is reachable under charge-balanced inputs if the system~\eqref{lifted_repetitive} is reachable. Further, the system~\eqref{main} is reachable under charge-balanced inputs if the system~\eqref{lifted_repetitive} is controllable. 

\subsection{Reachability and minimum-energy control}

At the time of $k=bh$, we arrive at 
\begin{align*}
    &x_{bh} = \bar A^{b} x_{0} + H_b \bar B w, &\text{where } H_b \coloneqq  \sum\nolimits_{i=0}^{b-1} \bar A^{i}
\end{align*}
with $\bar A=A^h$ and $\bar B=SQ$. Since $\|U_p\|_2^2=b \|w\|_2^2$,  the minimum-energy control problem~\eqref{min_control} under identical charge-balanced blocks becomes
\begin{align}\label{min_control_lifted2}
    \min_{w} \hspace{20pt}&J:=b\|w\|^2_2, \nonumber\\
    \text{s.t.} \hspace{20pt}& x_{(p+1)h} = \bar A x_{ph} +\bar B w,\\
        &  x_0=x_0, x_{bh}=x_f\nonumber.
\end{align}

\begin{lem}\label{reachability_repetitive}
    For the system~\eqref{lifted_repetitive}, a target system $x_f\in\R^n$ is reachable from an initial state $x_0$ if  $x_f -\bar A^b x_0 \in \Span(H_b \bar B)$. Then,  the unique minimum-energy solution to the problem~\eqref{min_control_lifted2} is
    \begin{equation*}
       w^* = (H_b \bar B)^\top \big( H_b \bar B (H_b\bar B)^\top \big)^{\dagger} (x_f-\bar A^b x_0).
    \end{equation*}
\end{lem}

Then, for the original system~\eqref{main}, $x_f$ can be reached from $x_0$ by applying $U^*=Qw^*$ to each input block, i.e., $U_0=U_1=\dots=U_{b-1}=U$.

Furthermore, if the matrix $H_b\bar B$ satisfies $\Rank(H_b\bar B)=n$, the system~\eqref{lifted_repetitive} is controllable for fixed $w$. Then, the original system~\eqref{main} is also controllable under charge-balanced inputs. Similar to the non-repetitive situation, we next investigate the conditions for $\Rank(H_b\bar B)=n$.

\subsection{Controllability}

The following lemma provides a sufficient condition to simplify the rank test $\Rank(H_b\bar B)=n$. 
\begin{lem}\label{full_H_p}
    For any  $b,h\in \BN$, it holds that $\Rank(H_b\bar B)=\Rank(\bar B)$ if there is no eigenvalue $\lambda$ of $A$ such that
    \begin{align}\label{condition_full_H_p}
        \lambda^{hb} =1 \text{  and } \lambda^h \neq 1. 
    \end{align}
\end{lem}

The proof of Lemma~\ref{full_H_p} follows from the lemma below, as $\Rank(H_b\bar B)=\Rank(\bar B)$ if $H_b$ has full rank.

\begin{lem}
    For any  $b,h\in \BN$, the matrix $H_b$ is invertible if and only if there is no eigenvalue $\lambda$ of $A$ such that
    \begin{align*}
        \lambda^{hb} =1 \text{  and } \lambda^h \neq 1. 
    \end{align*}
\end{lem}
\begin{pf}
    Define the polynomial
    \begin{equation*}
        f_b(z):= 1+z+z^2+\cdots+z_{b-1}.
    \end{equation*}
    Then, $H_b$ can be written as $H_b=f_b(\bar A)$. Let $\mu$ be an eigenvalue of $\bar A$ with corresponding eigenvector $v$, i.e., $\bar A v =\mu v $. Then, 
    \begin{equation*}
        H_b v = f_b(\bar A) v = \sum_{i=1}^{b-1} \bar A^i v = \sum_{i=1}^{b-1} \mu^i v = f_b(\mu) v. 
    \end{equation*}
    Thus, every eigenvalue of $\bar A$ produces an eigenvalue $f_b(\mu)$ of $H_b$. One can observe that $H_b$ is invertible if and only if $f_b(\mu)\neq 0$ for all $\mu \in \sigma(\bar A)$. 

    If $\mu\neq 1$, we have $f_b(\mu)= \sum_{i=1}^{b-1}\mu^i=\frac{1-\mu^b}{1-\mu}$, which implies that $f_b(\mu)=0$ if and only if $\mu^b=1$. If $\mu=1$, then $f_b(1) = b \neq 0$. Therefore, $f_b(\mu)=0$ if and only if no $\mu$ is a non-trivial $b$-th root of unity, i.e., for all $\mu\in\sigma(\bar A)$,
    \begin{align*}
        &\mu \neq \exp{\frac{2\pi k\eta}{b}}, &\forall k =1, \dots, b-1.
    \end{align*}
    Let $\lambda$ be an eigenvalue of $A$. As $\bar A=A^h$ be definition, it holds that $\mu=\lambda^h$ for all $\lambda\in\sigma(A)$. Observing that
    \begin{align*}
        \mu^b=1, \mu\neq 1 \Longleftrightarrow \lambda^{hb} =1, \lambda^h \neq 1
    \end{align*}
    completes the proof. \SQ
\end{pf}

Under the condition~\eqref{condition_full_H_p} in Lemma~\ref{full_H_p}, one only needs to check $\Rank(\bar B)$ to test the controllability. Observe that the dimension of $H_b\bar B$ is $n \times m(h-1)$. Therefore, a \textit{necessary condition} for $\Rank(\bar B) =n$ is $m(h-1)\ge n$. Next, we provide a sufficient condition for controllability. 

\begin{thm}\label{contro_rep}
    Let $h=2$. Assume the following conditions are satisfied:  (i) there is no $\lambda\in\sigma(A)$ such that $\lambda^{2b} = 1$ and $\lambda^2 \neq 1$, (ii) $1\notin \sigma (A)$, and (iii) $\Rank(B)=n$. Then, $\Rank(H_p\bar B)=n$, which means that the system~\eqref{main} is controllable under charge-balanced inputs. 
\end{thm}
\begin{pf}
    From Lemma~\ref{full_H_p}, the condition (i) implies that $\Rank(H_b\bar B)=\Rank(\bar B)$.     
    Next, we show $\Rank(\bar B) = n$. As $h=2$, we have 
    \begin{align*}
        &S=[AB \; B], &R=[I_m \; I_m].
    \end{align*}
    Without loss of generality, we select 
    \begin{equation}\label{selection_Q}
        Q=\frac{1}{\sqrt{2}}\begin{bmatrix}
            I_m\\-I_m
        \end{bmatrix}.
    \end{equation}
    Subsequently, one  can derive that 
    \begin{equation*}
        \bar B=SQ=\frac{1}{\sqrt{2}}(AB-B)=\frac{1}{\sqrt{2}}(A-I)B.
    \end{equation*}
    It follows from the assumption $1\notin\sigma(A)$ that $\Rank(A-I)=n$. Therefore, $\Rank((A-I)B)=n$ if $\Rank(B)=n$, which completes the proof. \SQ
\end{pf}

\begin{figure}[t]
	\centering
	\includegraphics[scale=0.6]{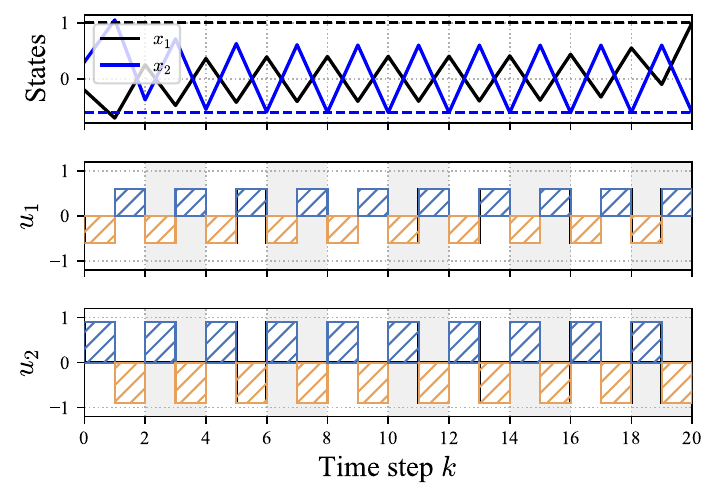}
	\vspace{-8pt}
	\caption{The minimum-energy inputs steer the system to a target state (dashed lines) in $N=20$ steps that consists of $b=10$ repetitive charge-balanced blocks.}	
	\label{repetitive_minimum}
\end{figure}

\begin{exmp}
    Consider the system matrices
    \begin{align*}
        &A=\begin{bmatrix}
           2 \;\;& 1 \\
            0 \;\;& 0.5
        \end{bmatrix}, & B=\begin{bmatrix}
            1&0\\
            0&1
        \end{bmatrix}.
    \end{align*}
    According to Theorem~\ref{contro_rep}, select $h=2$. Suppose that we want to steer the system from $x_0=[-0.2\;0.3]^\top$ to $x_f=[1\;-0.6]^\top$ in $b=10$ charge-balanced blocks. Next, we check if the conditions in this theorem are satisfied. First, the eigenvalues of $A$ are $2$ and $0.5$, satisfying Condition (ii). Further, (i) can be verified as no eigenvalue of $A$ is such that $\lambda^{20}=1$. Condition (iii) is naturally satisfied. By calculation, by calculation one can confirm that $\Rank(H_b\bar B)=2$. The minimum-energy control inputs can be computed following Lemma~\ref{reachability_repetitive}, which successfully drive the system to the target state as shown in Fig.~\ref{repetitive_minimum}. \QEDA
\end{exmp}

Notice that the condition (i) depends on $b$, indicating that one can carefully select $b$ to ensure it is satisfied.  It is worth noting that the conditions established in this theorem are still conservative. In particular, the condition $\Rank(B)=n$ implies $m\ge n$, which corresponds to a fully actuated system. This requirement is not needed in the non-repetitive setting. Next, we present an example in which this rank condition is violated, yet the system remains controllable under charge-balanced inputs. This motivates the development of additional, less restrictive conditions in the future.

\begin{figure}[t]
	\centering
	\includegraphics[scale=0.6]{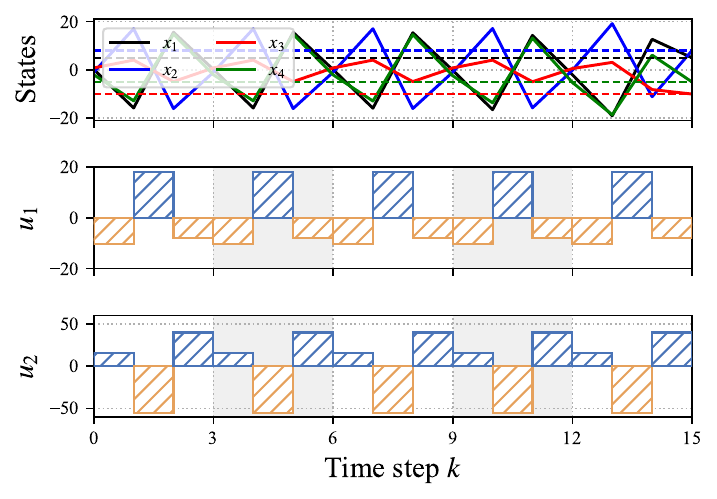}
	\vspace{-8pt}
	\caption{The minimum-energy inputs steer the system to a target state (dashed lines) in $N=15$ steps that consists of $b=5$ repetitive charge-balanced blocks.}	
	\label{repetitive_minimum_violating}
\end{figure}

\begin{exmp}
    Consider the system matrices
    \begin{align*}
        &A=\begin{bmatrix}
           1 \;& 2 \;&-2 \;& 1\\
           1 \;& 2 \;&2 \;& -1\\
           -1 \;& 1 \;& 3 \;& 1\\
           -6 \;& 6 \;&-6 \;& 8
        \end{bmatrix}, & B=\begin{bmatrix}
            1\;&0\\
            0\;&1\\
            1 \;&1\\
            0\;&1
        \end{bmatrix}.
    \end{align*}
     Observe that $\Rank(B)=2$, violating the condition in Theorem~\ref{contro_rep}. We choose $h=3$, indicating that $R=[I_2\;\;I_2\;\;I_2]$. Then, $Q$ can be chosen to be $Q=V\otimes I_2$ with
    \begin{equation*}
        V= \begin{bmatrix}
           1/\sqrt{2} \; \;& 1/\sqrt{6} \\
           -1/\sqrt{2} \; \;& 1/\sqrt{6} \\
           0 \; \;& -2/\sqrt{6}
        \end{bmatrix}.
    \end{equation*}
    Select $b=5$, and it can be demonstrated that $\Rank(H_b\bar B)=4$. Then, the system is controllable. As shown in Fig.~\ref{repetitive_minimum_violating}, the minimum-energy inputs calculated by following Lemma~\ref{reachability_repetitive} effectively steer the system to a desired state. \QEDA
\end{exmp}

\section{Conclusion}

This paper investigated the reachability and controllability of discrete-time linear systems subject to charge-balanced input constraints, motivated by electrical stimulation applications. We derived new sufficient conditions that characterize when charge balance limits controllability and when full-state steering can still be achieved. Minimum-energy charge-balanced control laws were obtained in closed form and validated through numerical examples. These results provide a systematic foundation for control design in systems that must satisfy charge-neutral actuation, with potential implications for developing more precise and adaptive brain stimulation strategies.

Several directions remain open. Some of the controllability conditions presented here are conservative, and relaxing these restrictions would be interesting for future research. Extending the framework to continuous-time models, nonlinear dynamics, and feedback controller design would further broaden its applicability to neurostimulation.

\bibliography{ifacconf}             

\begin{thebibliography}{16}
\providecommand{\natexlab}[1]{#1}
\providecommand{\url}[1]{\texttt{#1}}
\providecommand{\urlprefix}{URL }
\expandafter\ifx\csname urlstyle\endcsname\relax
  \providecommand{\doi}[1]{doi:\discretionary{}{}{}#1}\else
  \providecommand{\doi}{doi:\discretionary{}{}{}\begingroup \urlstyle{rm}\Url}\fi

\bibitem[{Bergey et~al.(2015)Bergey, Morrell, Mizrahi, Goldman, King-Stephens, Nair, Srinivasan, Jobst, Gross, Shields et~al.}]{bergey2015long}
Bergey, G.K., Morrell, M.J., Mizrahi, E.M., Goldman, A., King-Stephens, D., Nair, D., Srinivasan, S., Jobst, B., Gross, R.E., Shields, D.C., et~al. (2015).
\newblock Long-term treatment with responsive brain stimulation in adults with refractory partial seizures.
\newblock \emph{Neurology}, 84(8), 810--817.

\bibitem[{Dasanayake and Li(2012)}]{dasanayake2012charge}
Dasanayake, I. and Li, J.S. (2012).
\newblock Charge-balanced time-optimal control for spiking neuron oscillators.
\newblock In \emph{IEEE Conference on Decision and Control}, 1651--1656.

\bibitem[{Dasanayake and Li(2015)}]{dasanayake2015constrained}
Dasanayake, I.S. and Li, J.S. (2015).
\newblock Constrained charge-balanced minimum-power controls for spiking neuron oscillators.
\newblock \emph{Systems \& Control Letters}, 75, 124--130.

\bibitem[{Farooqi et~al.(2024)Farooqi, Vitek, and Sanabria}]{farooqi2024deep}
Farooqi, H., Vitek, J.L., and Sanabria, D.E. (2024).
\newblock Deep brain stimulation pulse sequences to optimally modulate frequency-specific neural activity.
\newblock \emph{Journal of Neural Engineering}, 21(3), 036045.

\bibitem[{Govindaraj et~al.(2023)Govindaraj, Paunonen, and Humaloja}]{govindaraj2023saturated}
Govindaraj, T., Paunonen, L., and Humaloja, J.P. (2023).
\newblock Saturated output regulation of distributed parameter systems with collocated actuators and sensors.
\newblock \emph{IFAC-PapersOnLine}, 56(2), 8940--8945.

\bibitem[{Kailath(1980)}]{kailath1980linear}
Kailath, T. (1980).
\newblock \emph{Linear Systems}.
\newblock Prentice-Hall Englewood Cliffs, NJ.

\bibitem[{Krauss et~al.(2021)Krauss, Lipsman, Aziz, Boutet, Brown, Chang, Davidson, Grill, Hariz, Horn et~al.}]{krauss2021technology}
Krauss, J.K., Lipsman, N., Aziz, T., Boutet, A., Brown, P., Chang, J.W., Davidson, B., Grill, W.M., Hariz, M.I., Horn, A., et~al. (2021).
\newblock Technology of deep brain stimulation: Current status and future directions.
\newblock \emph{Nature Reviews Neurology}, 17(2), 75--87.

\bibitem[{Liu et~al.(2024)Liu, Shi, and De~Schutter}]{liu2024stability}
Liu, C., Shi, S., and De~Schutter, B. (2024).
\newblock Stability and performance analysis of model predictive control of uncertain linear systems.
\newblock In \emph{IEEE Conference on Decision and Control}, 7356--7362.

\bibitem[{Mau and Rosenblum(2022)}]{mau2022optimizing}
Mau, E.T. and Rosenblum, M. (2022).
\newblock Optimizing charge-balanced pulse stimulation for desynchronization.
\newblock \emph{Chaos: An Interdisciplinary Journal of Nonlinear Science}, 32(1).

\bibitem[{Moehlis et~al.(2025)Moehlis, Zimet, and Rajabi}]{moehlis2025nearly}
Moehlis, J., Zimet, M., and Rajabi, F. (2025).
\newblock Nearly optimal chaotic desynchronization of neural oscillators.
\newblock \emph{arXiv preprint arXiv:2509.19531}.

\bibitem[{Ng et~al.(2024)Ng, Bush, Vissani, McIntyre, and Richardson}]{ng2024biophysical}
Ng, P.R., Bush, A., Vissani, M., McIntyre, C.C., and Richardson, R.M. (2024).
\newblock Biophysical principles and computational modeling of deep brain stimulation.
\newblock \emph{Neuromodulation: Technology at the Neural Interface}, 27(3), 422--439.

\bibitem[{Nonhoff et~al.(2023)Nonhoff, K{\"o}hler, and M{\"u}ller}]{nonhoff2023online}
Nonhoff, M., K{\"o}hler, J., and M{\"u}ller, M.A. (2023).
\newblock Online convex optimization for constrained control of linear systems using a reference governor.
\newblock \emph{IFAC-PapersOnLine}, 56(2), 2570--2575.

\bibitem[{Olumuyiwa and Kumar(2025)}]{olumuyiwa2025proportional}
Olumuyiwa, A.V. and Kumar, G. (2025).
\newblock Proportional-integral controller-based deep brain stimulation strategy for controlling excitatory-inhibitory network synchronization.
\newblock In \emph{American Control Conference}, 4627--4634.

\bibitem[{Porcari et~al.(2024)Porcari, Breschi, Zaccarian, and Formentin}]{porcari2024data}
Porcari, F., Breschi, V., Zaccarian, L., and Formentin, S. (2024).
\newblock Data-driven control of input saturated systems: a {LMI}-based approach.
\newblock \emph{IFAC-PapersOnLine}, 58(15), 205--210.

\bibitem[{Qin et~al.(2025)Qin, Pasqualetti, Bassett, and van Gerven}]{qin2025vibrational}
Qin, Y., Pasqualetti, F., Bassett, D.S., and van Gerven, M. (2025).
\newblock Vibrational control of complex networks.
\newblock \emph{IEEE Transactions on Control of Network Systems}.

\bibitem[{Tamekue et~al.(2025)Tamekue, Chen, and Ching}]{tamekue2025control}
Tamekue, C., Chen, R., and Ching, S. (2025).
\newblock On the control of recurrent neural networks using constant inputs.
\newblock \emph{IEEE Transactions on Automatic Control}.

\end{thebibliography}

\end{document}